\documentclass[prl,aps,twocolumn,groupedaddress,floats,showpacs]{revtex4}
\usepackage{graphicx}
\usepackage{dcolumn}
\usepackage{color}
\usepackage{bm}
\def\he4{$^4$He}
\def\Am3{\AA$^{-3}$}
\begin{document}

\author{L. Pollet}
\affiliation{Theoretische Physik, ETH Z\"urich, 8093 Z\"urich,
Switzerland}

\author{M. Boninsegni}
\affiliation{Department of Physics, University of Alberta,
Edmonton, Alberta T6G 2J1} \affiliation{BEC-INFM, Dipartimento di
Fisica, Universita' degli studi di Trento, Via Sommarive 14, 38050 Povo, Italy}

\author{A.B. Kuklov}
\affiliation{Department of Engineering Science and Physics, CUNY,
Staten Island, NY 10314}

\author{N.V. Prokof'ev}
\affiliation{Department of Physics, University of Massachusetts,
Amherst, MA 01003, USA} \affiliation{Russian Research Center
``Kurchatov Institute'', 123182 Moscow, Russia}

\author{B.V. Svistunov}
\affiliation{Department of Physics, University of Massachusetts,
Amherst, MA 01003, USA} \affiliation{Russian Research Center
``Kurchatov Institute'', 123182 Moscow, Russia}

\author{M. Troyer}
\affiliation{Theoretische Physik, ETH Z\"urich, 8093 Z\"urich,
Switzerland}
\title{Superfluidity of Grain Boundaries in Solid $^4$He}

\begin{abstract}
By large-scale quantum Monte Carlo simulations we show that grain boundaries in \he4 crystals are generically superfluid at low temperature, with a transition temperature of the order of $\sim$ 0.5K at the melting pressure; non-superfluid 
grain boundaries are found only for special orientations of the
grains. We also find that close vicinity to the melting line is not a necessary condition for superfluid grain boundaries , and a grain boundary in direct contact with the superfluid liquid at the melting curve is found to be mechanically stable and the grain boundary superfluidity observed by Sasaki {\it et al.} [Science {\bf 313}, 1098 (2006)] is not just a crack filled with superfluid.


\end{abstract}

 \maketitle

Superfluid grain boundaries (GB) were proposed as a plausible
scenario \cite{theorem,burovski,prelim} to explain the effect of
non-classical rotational inertia (NCRI) in solid \he4 discovered
by Kim and Chan \cite{KC}. An observation by Rittner and Reppy
\cite{Rittner} that NCRI signal can be eliminated through
annealing was the first explicit evidence that crystalline defects
are of crucial importance. The remarkable direct experimental
observation of grain boundary superfluidity (at the melting point) by Sasaki
{\it et al.} \cite{Sasaki}
confirms the early theoretical prediction and marks the beginning
of a new stage in the study of the supersolid phase of helium.

\begin{figure}[tbp]
\centerline{\includegraphics[width=0.7\columnwidth ]{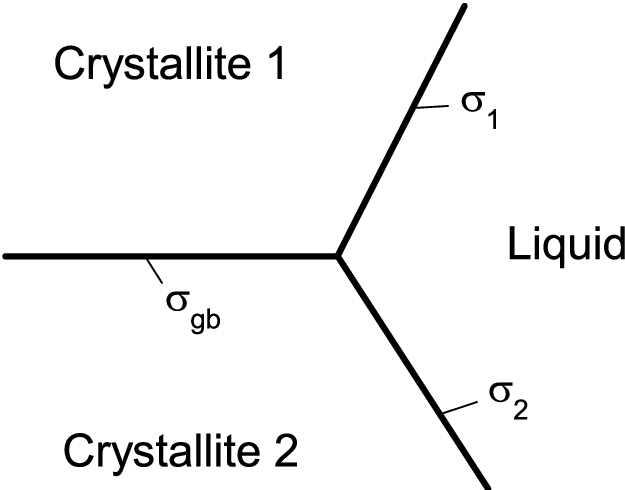}} \caption{Sketch of the equilibrium configuration of two crystallites in contact with liquid.  The configuration is mechanically stable only if the condition
(\ref{stab}) is met. Note that surface tensions $\sigma_1$ and
$\sigma_2$ need not be equal, because of the different
orientations of the crystalline axes.} \label{f1}
\end{figure}

In this Letter we expand on our previously reported preliminary results \cite{prelim} and show that a grain boundary in solid Helium is generically superfluid at low temperatures. The
transition temperature, $T_c$, is strongly dependent on the
crystallite orientation: while it is typically of the order of $\sim
0.5 \,$K, grain boundaries with special relative orientations of the two grains are found to be insulating (non-superfluid). 
We also obtain strong evidence that a grain boundary in contact with the superfluid liquid, as shown in Fig.~\ref{f1}, is mechanically stable.

This latter question is important for the interpretation of the experimental observation of superflow in crystals with grain boundaries \cite{Sasaki}. Since these experiments are carried out under the conditions of phase coexistence between a crystal and a liquid, the observed effect could either be true superflow along a grain boundary,  or rather  a thin liquid-filled crack in a crystal. The answer depends on the relationship between the three surface
tensions, $\sigma_1$, $\sigma_2$, and $\sigma_{\rm gb}$ (see Fig.
\ref{f1}). Mechanical and energetic stability of the
grain boundary require that
\begin{equation}
\sigma_{\rm gb}\; <\; \sigma_1\, +\, \sigma_2 \; . \label{stab}
\end{equation}
If this inequality is not satisfied, the liquid penetrates between
the two crystallites, and a crack is formed. We will show that this configuration is stable. The grain boundary superfluidity observed by Sasaki {\it et al.} is thus not merely the manifestation of a crack in the crystal filled with liquid.

Our  Path Integral Monte Carlo (PIMC) simulations are based on the continuous-space worm
algorithm~\cite{worm}. For spatial imaging, we employ two slightly
different techniques. The first consists of  producing {\it
condensate maps}, which are maps of the condensate
wave function, in thin slices of our sample. Within the
worm algorithm, this is accomplished by recording
spatial positions of the two open ends of the worms,
when they are sufficiently far away from  each other  such that correlations between them are
negligible~\cite{superglass}. As a result, the density of points
in the map is proportional to the condensate wave function. The
second technique is based on the {\it winding-cycle maps}. Here,
we collect statistics of instantaneous particle positions, by
considering only particles which participate in macroscopic
exchange cycles characterized by non-zero winding numbers, i.e.
cycles that directly contribute to the superfluid response~\cite{pollock87}.

\begin{figure}[tb]
\centerline{\includegraphics[scale=0.4]{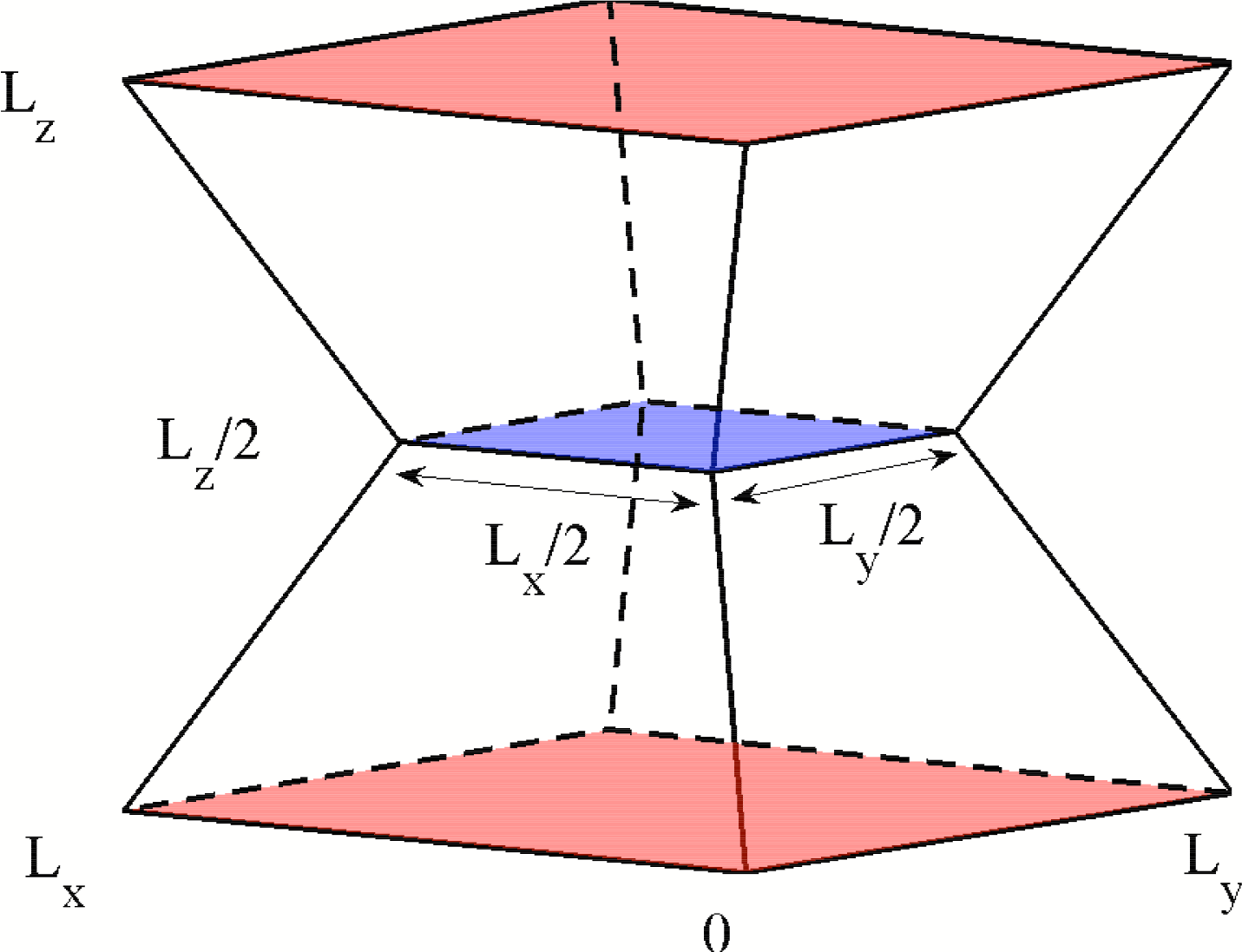}}
\caption{ (Color online) Initial setup for the simulation of a grain boundary in contact with a liquid. Two  truncated pyramids are placed on top of each other. 
The basal plane of both pyramids is a square with size $L_x \times L_y = L \times L$, with $L=24$ \cite{units}. The upper and lower pyramid have different random orientations and the height of both pyramids is $L_z = L/2$. The upper facets of the truncated pyramids are squares of initial size $L/2 \times L/2$, and form a grain boundary between the two crystallites, indicated by the blue square.  Liquid fills the volume outside the crystallites. 
Periodic boundaries are used in the $x$ and $y$-directions, while atoms in the $z=0$ and $z=L$ plane, drawn in red, are pinned, to prevent flow along the boundary in the $z$-direction. We refer to the Appendix for details such as the exact positions and rotation angles of the crystalline lattices within the pyramids.
\label{fig:pyrsetup}}
\end{figure}

{\it Stability of the grain-boundary--liquid junction}. In order
to check whether a grain boundary is destroyed when brought into contact with the
liquid, we perform a direct simulation of two crystallites in contact with liquid sketched in Fig.~\ref{f1}.
Our simulation setup, shown in Fig.~\ref{fig:pyrsetup}, consists of two truncated solid pyramids with random crystalline orientation placed on top of each other. The rest of the volume is filled with liquid. 

\begin{figure}[tbp]
\centerline{\includegraphics [bb=-180 75 592 430,
width=2.2\columnwidth ]{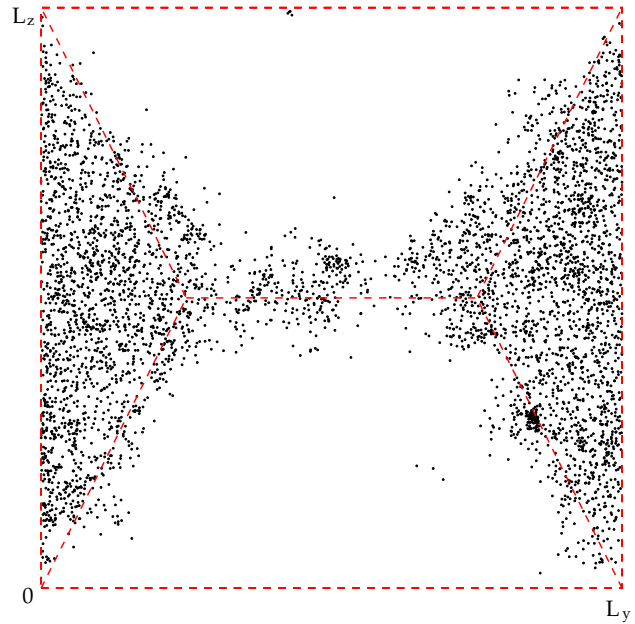}} \caption{Condensate map of the
two-pyramid sample. The map represents (by the density of points)
the condensate wavefunction in the slice $x\in [0.4L,\, 0.6L]$
averaged over the $x$-direction. The initial setup is shown by dashed lines.
} \label{f2}
\end{figure}

Since the goal of this
particular simulation is to study the stability of a grain boundary we can work at a relatively high temperature, $T$=0.8 K, which significantly enhances the performance of the
Monte Carlo simulations~\cite{RMB}. 
The simulation is carried out in the grand canonical ensemble,
with the chemical potential fixed at the phase coexistence point. The
equilibrium number of atoms in our sample was measured to be about
$13660$.

In order to stabilize the solid phase in the system, we pin (i.e., do not update) solid
atoms in the vicinity of the pyramid bases. 

In Fig.~\ref{f2} we show the condensate map of the sample. We see
that the grain boundary between the two crystallites is a robust quantum object 
with a thickness of order 3 (see also Fig.~\ref{f4})
and did not disappear during the simulation run. The system has converged to a state where the liquid and solid phase coexist. Compared to the initial configuration, the shapes of the crystallites (including the angles) have noticeably changed. This is not surprising since the optimal shape of the crystal-liquid interface depends on the particular orientation of the crystallite axes with respect to the interface.

{\it Superfluidity of grain boundaries}. Having confirmed the stability and superfluidity of grain boundaries in a generic sample, we next perform a more systematic study of the properties of grain boundaries, extending our previously reported simulations~\cite{prelim}, 
where  the superfluidity had been
observed, but only for one particular polycrystalline structure,
and grain boundary superfluidity could not clearly be
distinguished from  grain edge superfluidity.

The simulation of a grain boundary bewteen two truncated pyramids with basal length $L=24$ is computationally very expensive since it requires $\sim$ 13660 atoms,  and is by far the largest simulation of solid Helium performed to date. 
Reducing the size of the pyramids is not an option,
since the typical width of the superfluid grain
boundary region, estimated from Fig.~\ref{f2} (see also Fig.~\ref{f4}),
is $\sim 3$.  Hence, in the two-pyramid samples of linear sizes
significantly smaller than $L=20\div 30$, the grain boundary is not well defined, and the system is not supposed to be metastable. For instance, in a sample with $L=12$, we found that the grain
boundary and the two crystallites (apart from the pinned
atoms) melt.

\begin{figure}[tb]
\centerline{\includegraphics[scale=0.4]{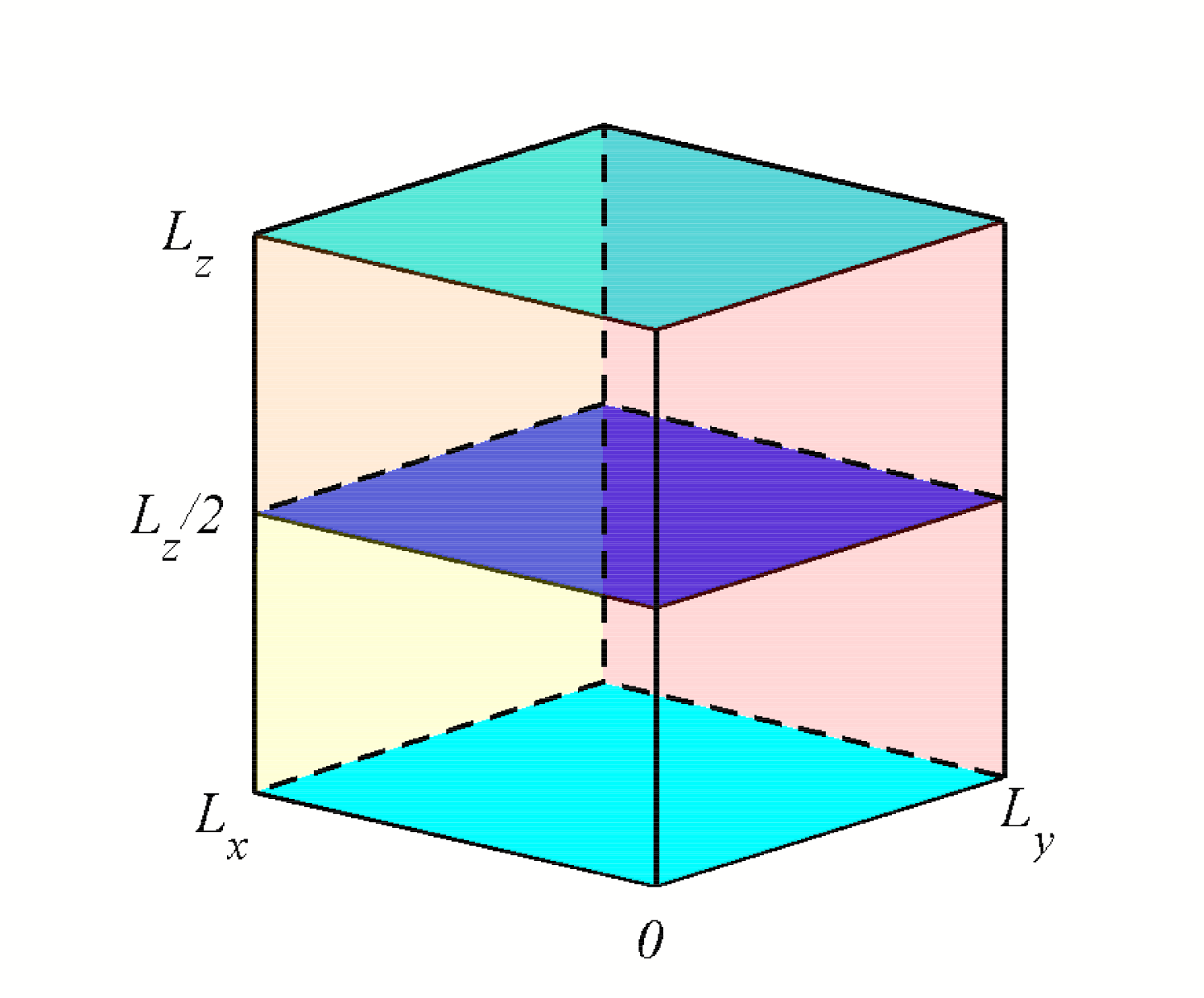}}
\caption{(Color online) Sketch of the initial of two cuboid crystallites with a total of about 2000 atoms placed on top of each other. 
The basal plane of both cuboids is a square with size $L_x \times L_y = L \times L$, with $L=12$. The upper and lower cuboid have different random orientations (see the Appendix for more details). The height of both cuboids is $Lz/2 = 7$, yielding a grain boundary in the $z=0$ plane (light blue) and a grain boundary in the $z=Lz/2$ plane (dark blue). Periodic
boundary conditions are used along all directions, but motion across the $xz$ boundary was suppressed by pinning all
atoms at a distance smaller than 0.75 from the $y=0~(y=L_y)$ boundary indicated in red. This ensures that any superfluid response in the
$x$-direction is due to the superfluidity of two horizontal grain boundaries\cite{shunt}. 
An additional grain boundary due to periodic boundary conditions arises for each crystallite at the $x=0~(x=L_x)$ plane (yellow and orange for the lower and the upper cuboid, respectively), but
these do not affect the superfluid response in the $x$-direction.
\label{fig:cuboidsetup}}
\end{figure}

To study the properties of generic grain boundaries we use the
different sample geometry shown in
Fig.~\ref{fig:cuboidsetup} consisting of 2000 atoms . Two cuboids of equal size and with
random crystalline-lattice orientations are placed on top of each
other along the $z$-direction, which creates two parallel grain
boundaries in the $xy$-plane due to periodic boundary conditions.

\begin{figure}[tbp]
\centerline{\includegraphics [bb=-150 460 530 620,
width=2.2\columnwidth ]{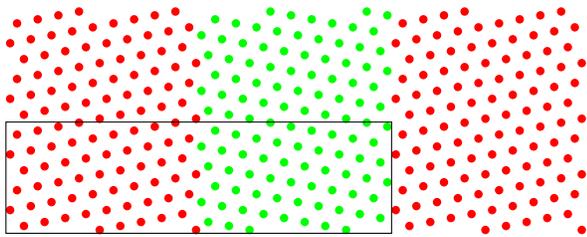}} \caption{Example of
an insulating grain boundary. This grain boundary is between two
crystallites that differ only by a rotation about the axis
perpendicular to the base plane. Shown in the figure is a base
plane. The rectangular box is the actual simulation cell, the rest
being a periodic continuation of the sample. Even with this
moderate system size, it is clearly established that the grain
boundary is not superfluid. Decreasing the angle between the
two crystallites strengthens the insulating character, due to the better match between the atoms of
different crystallites. In the small-angle limit, this grain
boundary can be viewed as an array of dislocations; the latter are
inevitably insulating, given the fact that even at moderate angles
the grain boundary is not superfluid.} \label{f3}
\end{figure}

An important result of our simulations is that, as already observed in our preliminary studies \cite{prelim}, {\it not all grain boundaries in the {\it hcp} \he4 crystal
are superfluid}.  Grain boundaries featuring an extra symmetry
such as stacking faults and special grain boundaries with nicely matching angles similar to the one shown in Fig.~\ref{f3}, are insulating.

\begin{figure}[tbp]
\centerline{\includegraphics [bb=-310 80 560 260,
width=5.\columnwidth ]{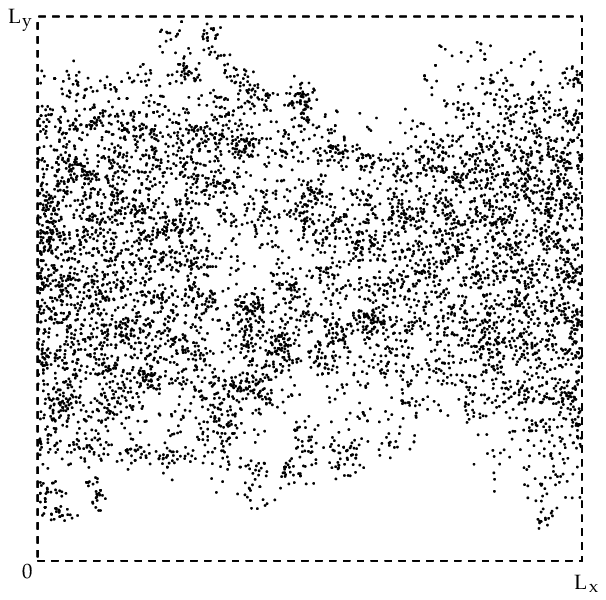}}
\centerline{\includegraphics [bb=-310 90 560 280,
width=5.\columnwidth ]{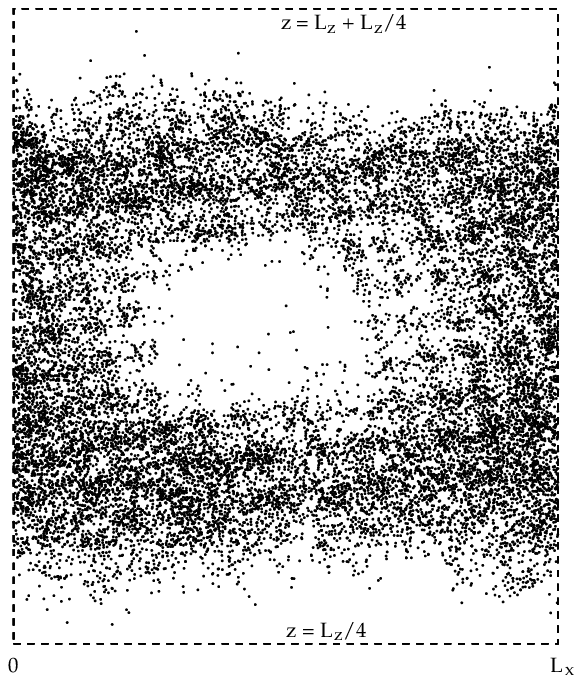}} \caption{Phase coherence
properties (winding-cycles maps) of grain boundaries in the \he4
sample shown in Fig.~\ref{fig:cuboidsetup}. Upper panel: projection on the $xy$ plane of the data
for the upper half of the sample containing one of the two grain
boundaries. Lower panel: ($z$-shifted) projection of all the data
points on the $xz$ plane. 
Note that the
$x=0$ ($x=L_x$) grain boundary (induced by boundary conditions) of
one of the two crystallites turned out to be insulating. This is
yet another example of insulating grain boundary with special orientation of
crystalline axes.} \label{f4}
\end{figure}

The generic case, however, are superfluid grain boundaries. Results of a typical simulation at
$T=0.25 \,$K and $n=0.0287\,$\Am3 (melting density) are shown in Fig.~\ref{f4} . The
winding-cycle map clearly reveals superfluidity along the
$x$-direction. Based on the relatively large system size utilized
in this study, we argue that our observed superfluid signal
reflects macroscopic grain boundary superfluidity, and is not an
artifact, due for example to the vicinity of a superfluid
ridge --- the intersection of the grain boundary of interest and the additional grain boundary caused by periodic boundary conditions
at $x=0$ ($x=L$). 
Incidentally, we note that in the area close to
the ridge, of size $\sim 3$, the density of the map is
significantly increased, which is consistent with our general
observation that nearly all ridges have robust phase-coherence
properties.

An interesting observation is that the density in the vicinity of superfluid
grain boundaries is close to that of the crystal, which again confirms that 
the superflow along the grain boundary is not a liquid-filled crack.

On the basis of a number of simulations similar to the one
presented in Fig.~\ref{f4}, performed at different pressures,
temperatures, and crystallite axes orientations, we conclude that
generic grain boundaries are superfluid with typical (orientation dependent)
transition temperatures  of about half a Kelvin. The width of the
superfluid grain boundary region is $\sim 3$. We conjecture that the maximum
possible $T_c$ for grain boundary should be at least smaller than the
transition temperature of the overpressurized liquid of the same
density as the crystal.  By simulating
superfluid properties of the overpressurized liquid, we found the
transition temperature at the density $n=0.0287\,$\Am3 to be
$1.5(1)\,$K. We thus take this value as the upper bound for
grain-boundary $T_c$ at the melting pressure.

In order to provide a further assessment of the robustness of our
conclusions, we have repeated the same study, replacing helium
with molecular {\it para}-hydrogen, at the same low temperatures.
In this case, individual particles have a mass which is half that
of helium atoms, whereas the interaction potential is
approximately three times deeper. No evidence of superfluidity was
ever observed in this case, in agreement with experimental findings~\cite{Chan_hydrogen}.

Summarizing, based on a direct quantum Monte Carlo simulation of a
grain boundary in {\it hcp} \he4 in contact with liquid under the
conditions of phase coexistence, we argue that it is
thermodynamically stable against dissolution  into two
crystallites separated by a crack. This lends theoretical support
to the observation of Sasaki {\it et al.} \cite{Sasaki} of generic
superfluid grain boundaries. We have studied superfluid properties of grain boundaries and
found that special grain boundaries of higher symmetry are insulators. On the
other hand, grain boundaries of a general form are found to be superfluid, with
typical transition temperatures of the order of $\sim 0.5$ K.

While these results lend support to an explanation of the results of Kim and Chan  \cite{KC} as grain boundary superflow, the direct relevance of these findings to the experiment by Kim and Chan is
open, until the presence or absence of grain boundaries is investigatred by X-ray or neutron scattering experiments on the torsional oscillator cells used in these experiments.

We thank S. Balibar, M. Chan, R. Hallock, and J. Reppy for
fruitful discussions. This work was supported by the Swiss National Science Foundation, as well as the US National
Science Foundation under Grants Nos. PHY-0426881, PHY-0426814 and
PHY-0456261, and by the Natural Science and Engineering Research
Council of Canada, under research grant 121210893. NP gratefully
acknowledges hospitality and support from the Pacific Institute of
Theoretical Physics, Vancouver (BC). Part of the simulations were
performed at the Hreidar cluster at ETH Zurich. LP , BS and MT acknowledge support of the Aspen Center for Physics.

\appendix*

\section{APPENDIX : PREPARATION OF THE INITIAL SAMPLE}

\subsection{Euler angles}

The rotation from the $x,y,z$ frame to the $x'y'z'$-frame is completely determined
by the three Euler's angles $(\theta,\phi,\psi)$. The first rotation is by an angle $\phi$ about the $z-$axis, the second is by an angle $\theta \in [0, \pi]$ about the $x-$axis, and the third by an angle $\psi$ about the $z-$axis (again).
The transformation matrix ${\bf U}$, defined as ${\bf r}={\bf r}_0+{\bf U}\, {\bf r}'$, reads:
\begin{eqnarray}
U = \left( 
\begin{array}{ccc}
c \psi c  \phi -c  \theta  s  \psi s  \phi & c \psi s \phi +c \theta s \psi c \phi & s \theta s \psi \\
-s \psi c \phi -c \theta c \psi s \phi & -s \psi s \phi +c \theta c \psi c \phi & s \theta c \psi \\
s \theta s \phi & -s \theta c \phi & c \theta
\end{array}
\right),
\end{eqnarray}
where $c$ is a shorthand notation for $\cos$ and $s$ for $\sin$.

\subsection{Construction of the initial sample with the pyramids, see Fig.~\ref{fig:pyrsetup}}\label{sec:pyr}
Based on the crystallographic definition of
the $x'y'z'$-system, the position of the first atom, ${\bf
r}'=(0,0,0)$, unambiguously defines an infinite {\it hcp} lattice,
of which we take only the atoms whose positions are inside the
pyramid. 
The randomly selected values which specify the initial
configuration of the lower crystallite are shown in Table~\ref{table:lowerpyramid}.
\begin{table}[h]
\begin{tabular}{|l|r|}
\hline
parameter & value \\
\hline 
$x_0$ & 34.946 \\
$y_0$ & 24.507 \\
$z_0$ & 0.87081 \\
$\theta$ & 1.9978 \\
$\phi$ & 4.0671 \\
$\psi$ & 2.3174 \\
\hline
\end{tabular}
\caption{Initial parameters for the lower pyramid.} \label{table:lowerpyramid}
\end{table}

The second pyramid is constructed similarly, and then
mirrored with respect to the $z=L/2$ plane.  
The randomly selected values which specify the initial
configuration of the upper crystallite are shown in Table~\ref{table:upperpyramid}.

\begin{table}[h]
\begin{tabular}{|l|r|}
\hline
parameter & value \\
\hline 
$x_0$ & 3.9613 \\
$y_0$ & 37.245 \\
$z_0$ & 19.288 \\
$\theta$ & 0.98923 \\
$\phi$ & 1.0833 \\
$\psi$ & 1.9885 \\
\hline
\end{tabular}
\caption{Initial parameters for the upper pyramid.} \label{table:upperpyramid}
\end{table}

The initial positions of atoms of the liquid component are selected at random. The liquid fills the volume outside the crystallites.

\subsection{Construction of the initial sample with the cuboids, see Fig.~\ref{fig:cuboidsetup}}

The initial sample has about 2000 atoms consisting of  two cuboid crystallites placed on top of each other along the $z$-direction. The crystallites were prepared in the same way as the pyramids in Section~\ref{sec:pyr}: we choose a random position $\bm{r_0}$ and random Euler angles $(\theta, \phi, \psi)$ in order to specify the offset and the orientation of the crystallites versus the simulation box. There is no liquid phase in the intial sample. The values of the initial parameters of the lower cuboid are shown in Table~\ref{table:lowercuboid}.

\begin{table}[h]
\begin{tabular}{|l|r|}
\hline
parameter & value \\
\hline 
$x_0$ & 34.946 \\
$y_0$ & 24.507 \\
$z_0$ & 1.0160 \\
$\theta$ & 1.9978 \\
$\phi$ & 4.0671 \\
$\psi$ & 2.3174 \\
\hline
\end{tabular}
\caption{Initial parameters for the lower cuboid.} \label{table:lowercuboid}
\end{table}

The values of the initial parameters of the upper cuboid are shown in Table~\ref{table:uppercuboid}.

\begin{table}[h]
\begin{tabular}{|l|r|}
\hline
parameter & value \\
\hline 
$x_0$ & 6.6684 \\
$y_0$ & 28.603 \\
$z_0$ & 19.706  \\
$\theta$ & 2.1532  \\
$\phi$ & 4.7702 \\
$\psi$ & 6.0979 \\
\hline
\end{tabular}
\caption{Initial parameters for the upper cuboid.} \label{table:uppercuboid}
\end{table}


\begin{thebibliography}{99}
\bibitem{shunt} Otherwise, we would have to
deal with the ``shunting" superfluid response of grain boundaries arising, by
periodicity, within the same crystallite at $y=0~(y=L_y)$
interfaces.
\end{thebibliography}
\end{document}